\documentclass[12pt]{article}
\usepackage{graphics}
\usepackage[dvips]{graphicx}
\usepackage[left]{lineno}
\usepackage[square,sort,comma,numbers]{natbib}
\usepackage{url}
\begin{document}
{\footnotesize jcis@epacis.org}

\begin{center}

{\bf Xitris: A Software to Acquire, Display, Compress, and Publish Data in Real
Time using Distributed Mode for the Solar Radio Interferometer RIS}
\bigskip


{\small Victor H De la Luz$^{a}$\footnote{E-mail Corresponding Author:
    itztli@gmail.com} and Alejandro Lara$^{b}$}
\smallskip

{\small
$^a$Departamento de F\'isica, Universidad Aut\'onoma Metropolitana Iztapalapa, A. P. 55-534, C.P. 09340, D.F., M\'exico.\\

$^b$ 
Instituto de Geof\'isica, Universidad Nacional Aut\'onoma de M\'exico, 
M\'exico D.F., M\'exico
}
{\footnotesize Received on *** / accepted on *****, 2014}

\end{center}


\begin{abstract}
Upgrading the infrastructure of old scientific instruments requires   
the development of new hardware and software
which may be expensive (in general, 
%
these projects lack of enough resources to acquire fast and modern 
infrastructure 
to become competitive and functional in the era of digital data). 
Particularly, 
the development of software for data acquisition in real time is one of the most 
important topics in experimental science and industry. 
However, the constant improvements on the  data acquisition hardware, 
discourages the development of software highly optimized in order to minimize 
the consumption of  resources  like 
processor time and read/write memory, etc.
In this work, we present a Open Source code called ``X Interface to RIS'' 
(Xitris). 
This is a modular and distributed code which,
using relatively slow processors and low memory hardware,  
 is able to acquire, display, 
compress and publish (via Internet) digital data  in real time. 
Xitris was developed for the Solar Radio Interferometer (RIS) at Geophysics 
Institute of the National University of Mexico (UNAM) but nowadays is 
working in another 3 radio observatories.

\bigskip

{\footnotesize
{\bf Keywords}: distributive algorithms, analog/digital converter.}
\end{abstract}

\textbf{1. Introduction}

\bigskip
\bigskip

The Analog/Digital Converters (ADC) translate analog electrical signals that 
represents physical parameters (for example pressure, light, temperature) in 
digital data, represented in binary code. An ADC has four main characteristics: 
the velocity of the conversion (Hz), the resolution of the data (bits), 
the noise of the conversion (mV), and the range of conversion (V). 
The ADC hardware delivers the data sequentially. The process control, e. g. 
 acquisition rate, error control and storage, is controlled by software. 
Although 
nowadays, to make ADCs boards 
 is relatively cheap, 
the challenge remains in the software development.

The Laboratory of Solar Physics in the Institute of Geophysics of the
National University of Mexico is responsible of the Solar Radio Interferometer
(RIS, Figure 1). This instrument was donated by  the former Soviet Union 
30 years ago. 
At the beginning, the RIS was operated at the National Institute of 
Astrophysics (INAOE) in the state of Puebla, Mexico, and few years later 
 the RIS was moved to its actual location in Mexico City.

The RIS has been updated several times in the past 30 years, 
unfortunately, most of the 
documentation was lost or are in Russian language and obsolete, 
the available funds for the project
are very limited and the computer infrastructure 
at the time when we started the XITRIS project was old. At that time, the
laboratory bought a comercial Analog/Digital Converter (ADC) to improve the  
record of the RIS data.

The acquisition of a new ADC board implied the necessity of either buying or 
developing 
new software. 
The former option was avoided due to the limited budget and 
due to the fact that the laboratory is focused in the development of 
technology.

The 
attempt to create a new acquisition software using  LabVIEW
package failed \citep{travis2007labview}. 
An analysis 
showed that the old hardware infrastructure, with monolithic design, 
prevented the running of   
the ADC
acquisition software. 
\citep{nissanke1997realtime}.  
Therefore, the design of a new distributed software was necessary.


In this work, we present a general purpose algorithm intended 
to  acquire, display, compress, and publish data in real time and 
in a distributed mode. 
%
%
The ``X Interface to RIS'' (Xitris) code can be distributed in four
computers in order to maximize the use of resources, and may be used in any 
observatory with high rate of data acquisition.
%
The software engineering process was carried out using eXtreme Programming (XP).

\bigskip
\begin{center}
\includegraphics[height=8cm]{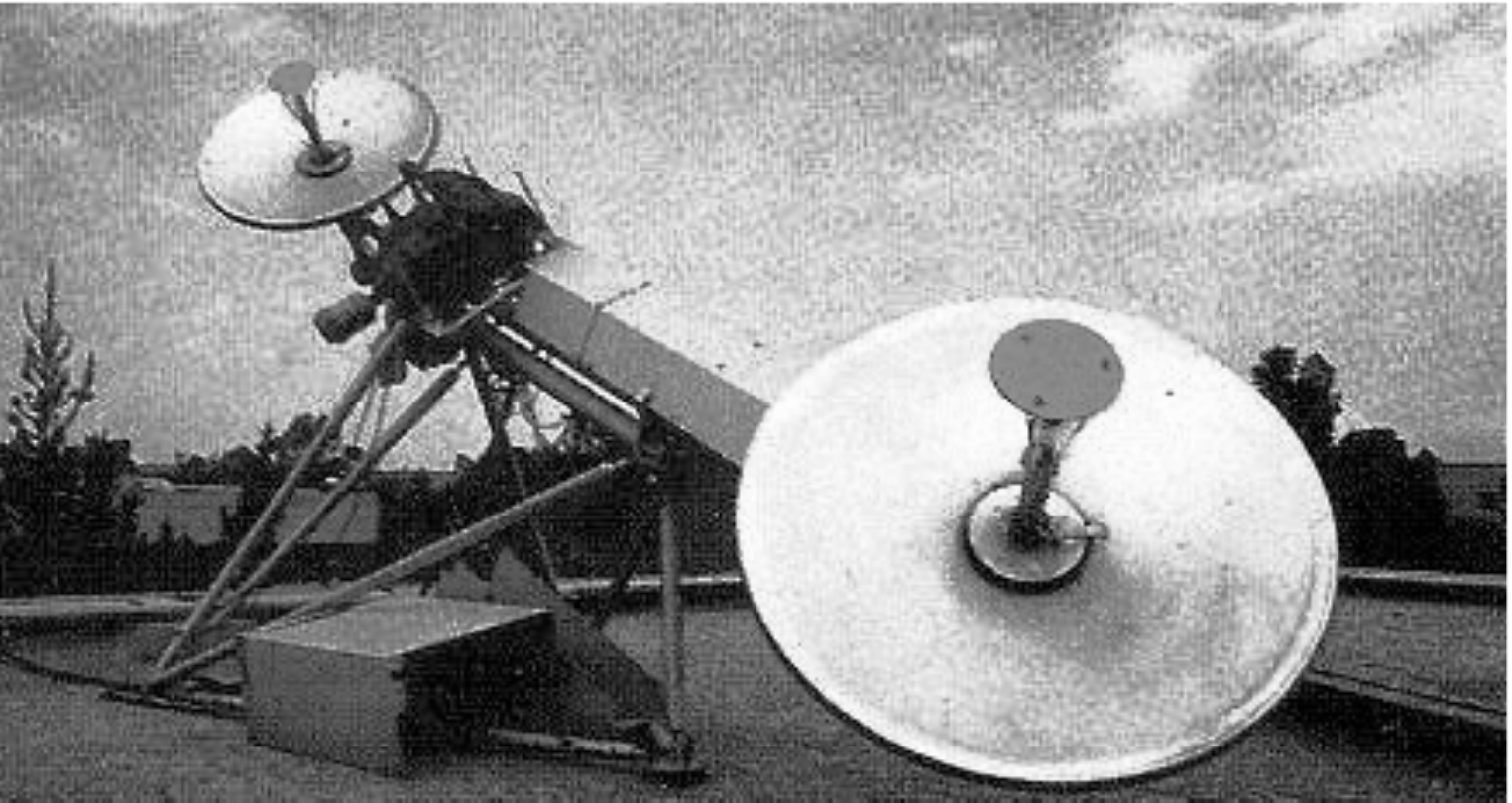}
\bigskip
\textbf{Figure 1} - The Solar Radio Interferometer (RIS) in its final 
location 
at the National University in Mexico, in Mexico City.
\bigskip
\end{center}

The paper is organized as follows: 
In  Section 2 we present the status of the old Recorder System and  
the infrastructure of the Laboratory at the time of the project starting, 
respectively. The 
algorithm and its implementation is presented in Section 3. 
In Section 4 we show the implementation results and
finally, in Section 5 are our conclusions.

\bigskip
\bigskip
\textbf{2. The Solar Radio Interferometer (RIS)}
\bigskip
\bigskip

The old recorder system of the RIS had four chart recorders  which 
printed a stroke in continuous paper representing four characteristics 
(total and polarized flux as well as two channels of interferometric signal)
of the solar 
flux at 7 GHz  
observed and processed by the RIS (Figure 2, shows an example of the total flux). 
This kind of plots of (counts versus time) is the main piece of work for the
researchers interested in solar activity. 

\bigskip
\begin{center}
\includegraphics[height=8cm]{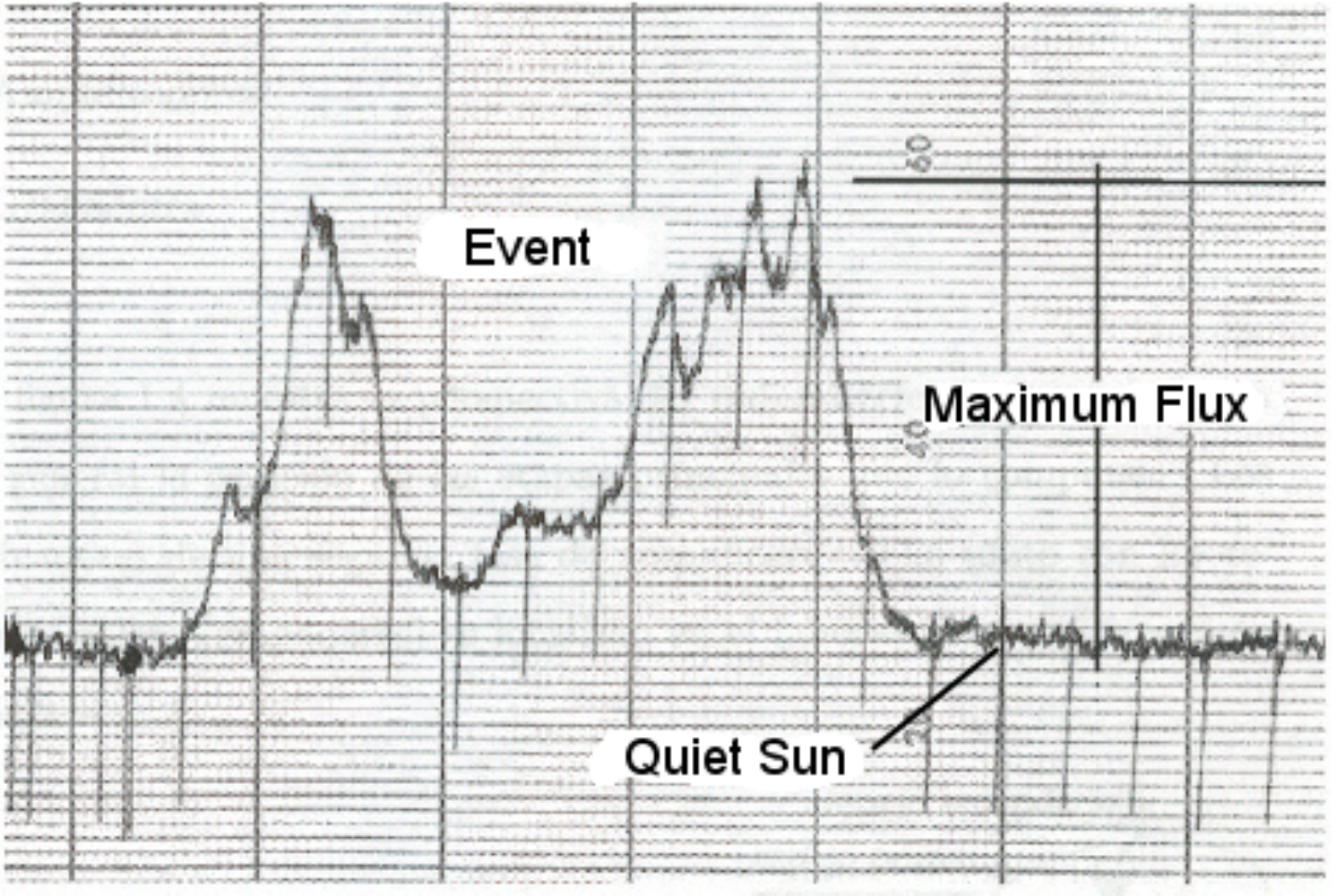}
\bigskip

\textbf{Figure 2} - One example of a solar flare observed in the total 
flux channel and recorded in continuous paper using the old recording system 
of the Solar Radio Interferometer.
\bigskip
\end{center}

The telescope started manually by the 
technical staff, 
they turned on
the solar radiotelescope every day
around the sunrise and marked (in the same paper of the signal) two  
reference  levels, one with no source (zero) at the input and other form a known 
noise source.  
Finally, 
the radiotelescope was manually pointed  to the Sun position, 
and the tracking system was started when
the signal of the total flux channel reached its maximum level. 

Once the tracking system was working, a regular line at a constant level 
was printed, 
marking the 
quiet sun signal. 
An abrupt increase of the signal may be caused by a solar event or artificial interference.

At the beginning of the project, the laboratory had the follow
infrastructure:
\begin{enumerate}
\item Network
  \begin{itemize}
  \item 2 IPs homologous address.
  \end{itemize}
\item Hardware
  \begin{itemize}
  \item 3 Computers Pentium MMX at 166MHz with 46 Mb RAM memory, with Bus
    Master at 66 MHz and video card Cirrus Logic GD 5430.
  \item Computer 80486 with ISA interface.
  \item Laser Printer.
  \item Digital converter board ADC LabPC+ with ISA interface.
  \end{itemize}
\item Software
  \begin{itemize}
  \item Debian GNU/Linux.
  \item Windows 95.
  \end{itemize}
\item Documentation
  \begin{itemize}
  \item Lab-PC+ User Manual.
  \end{itemize}
\end{enumerate}
The main problem was the old infrastructure and the limited
resources for the project. 
At the early stage of the project 
the
group focused in a specific purpose software intended to solve the 
problem of the old RIS
infrastructure.  However, we realize that more observatories and/or 
laboratories may be in a similar situation than the RIS (in terms of software 
requirements) and therefore, we  decided to build a general propose adquisition data 
software using the RIS has a study case.

\bigskip
\begin{center}
\includegraphics[width=14cm]{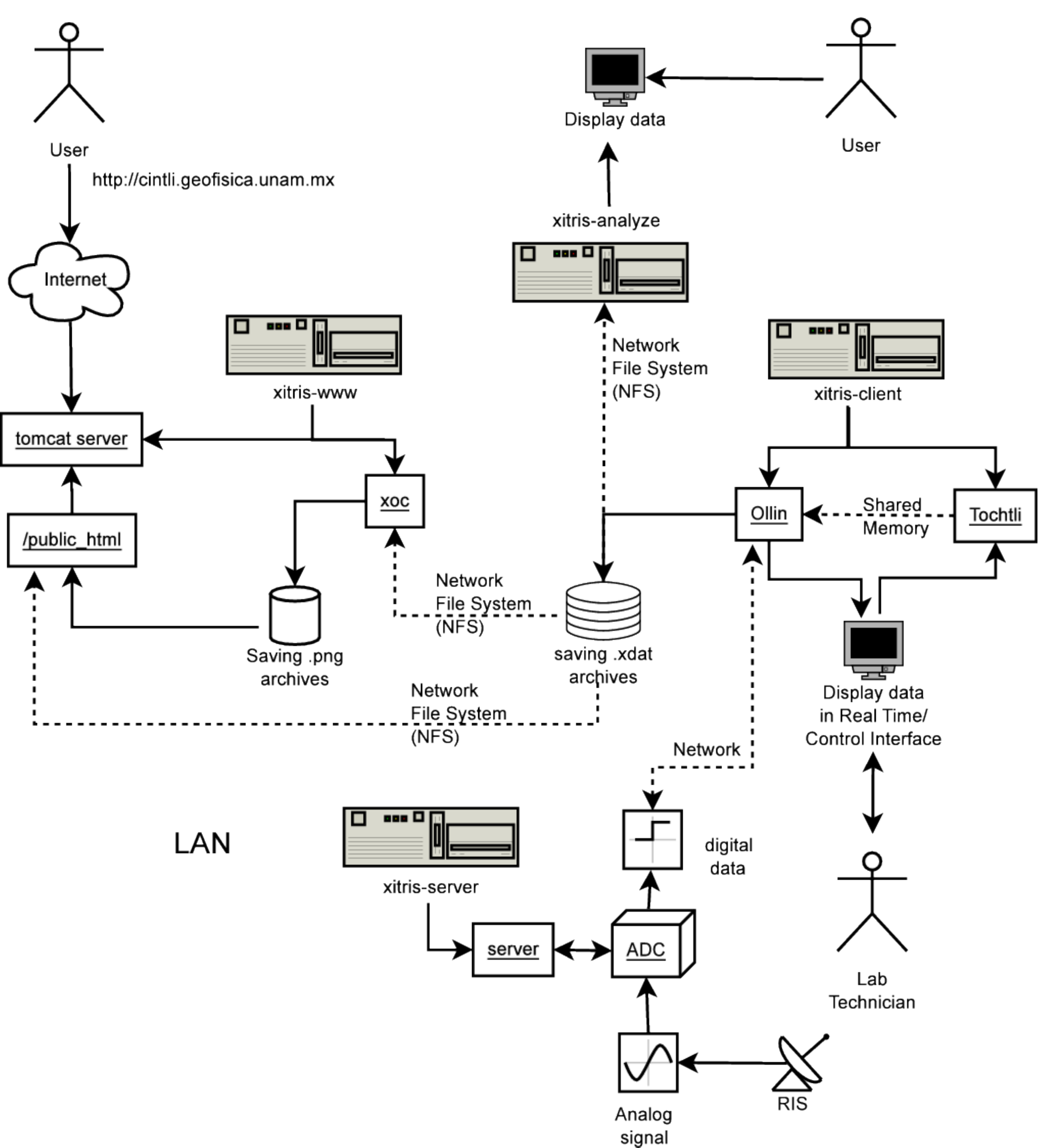}
\bigskip

\textbf{Figure 3} - Work-flow model for the Xitris project. 
The xitris-server collect the data and send the information by network.  
The Xitris-client saves and display the data. 
Xitris-www publish and display the data in the Internet. 
Xitris-analyze shows the data to other people (researchers) inside the LAN. 
\bigskip\label{figure3}
\end{center}

\bigskip
\bigskip
\textbf{3. The X Interface to the RIS}
\bigskip
\bigskip

The X Interface to the RIS (Xitris) is written in C using GTK+ and GDK
libraries \citep{harlow1999developing}, except the server
web-pages, for these cases we use Java \citep[JSPs technology
  see Section][]{hall2000core}. The code runs over GNU/Linux systems. We compile the
code with GCC compiler. We use eXtreme Programming (XP) as software
development methodology \citep{beck2000extreme}. The design includes the
client-server paradigm, 
distributed systems and modular object-oriented
technique (OOP) but using C language. 
It is interesting to note that although the C
language is not OOP (like C++ or Java) we can use the C native structures to
emulate the OOP in a very efficient way in order to encapsulate the
functionality of each piece of  software and build a modular system that
can be replace each library easily \citep{van1990data}. 

In our case,
the most important library is related with the ADC control board.
We developed generic methods which are able to 
be reprogrammed in order to include new ADC boards.

In eXtreme Programming there are several (short) cycles of development.
The scope of each cycle is focused in: 
writing a short version, using simple
design, testing, re-factoring, and pair programming. 
In this way, the process of
building the software is fast and dynamic. 

Prior of the starting of each cycle, it is necessary to carry out 
several interviews with the users in order to determine the scope of
the next step of development. 

The XP cycle have 4 steps:
Analysis, Design, Implementation, and Tests.
%
It is necessary
analyze each step of development, design the
solution, implement it, and finally test the code. 

The XITRIS software has 14 cycles of development:
\begin{enumerate}
\item Development of the Client-Server algorithm.
\item Integration of the Server with the LabPC+.
\item Development of the graphic interface.
\item Implementation of Start/End data acquisition.
\item Data plotting.
\item Saving data.
\item Database structure.
\item Saving the status of XITRIS.
\item Implementation of tools to change of Amplification and Frequency of the 
data  acquisition.
\item Publishing the data in Internet in Real Time.
\item Plotting the data in Real Time.
\item Implementation of the data display.
\item Detecting Solar Events.
\item Creation of install packages.
\end{enumerate}

The final design includes 4 general modules (see Figure 3):
\begin{enumerate}
\item The xitris-server: 
acquires data from an ADC board and send it  trough the
  network. This server also configures the ADC board.
\item The xitris-client: receives the data from the xitris server, 
and saves and displays it the data in real time. 
\item The xitris-www:  plots and publishes the data  on the internet in 
real time. This
  module also shows an interface of the database on the internet 
and produces dynamically the  web-pages with the data plots in real time.
\item The xitris-analytic: is intended to display, log, and analyze the data in real time.
\end{enumerate}
The xitris modules can be installed in a single or  multiple computer
environment. We design this 4 pieces of software in order to maximize the
computer infrastructure of the RIS laboratory.


In the following sub sections we will describes each module and their interactions between them.
\newpage
\bigskip
\bigskip
\textbf{3.1 xitris-server}
\bigskip
\bigskip

The xitris-server is a single piece of software that listen the calls from
the xitris-clients \citep[in the paradigm of client-server the xitris-server
  module play the role of server][]{kochan1989unix}.  
The remote call includes the configuration of the ADC board
(sampling frequency and amplification), the starting and the ending 
of the acquisition
process, and the error handling. For the case of the Lab-PC+, we can handle up
to 4 channels simultaneously with 12 bits of precision.

The server publish a port and wait for a client request. 
Then, negotiate the data transmission: First, the server receives 20 
integers (with the configuration from the client), configures the ADC, 
starts the data acquisition and sends, 
throughout the network socket, 4 integers each time (for each channel), if an 
error is detected in the ADC, the algorithm sends a signal to the client
reporting that an error was detected, the client handles the error, 
if a change of configuration is requested, the acquisition is stopped, the ADC 
is reconfigured and then the acquisition restarts. 

The server has modular architecture, the ADC library can be reprogrammed in
order to conserve the structure and include a new ADC board. For the case of
the RIS, we implement the Lab-PC+ library.

The server reads the 12 bits of data from the ADC and transform it to 
A2 complement integer of 32 bits (See Appendix A.1). This implementation is very fast and give us the possibility of read data in
the 4 channels in milliseconds with a very old computer.

The xitris-server was installed in a old 486 PC with the ADC board. 
We tested the acquisition time (without graphics or data saving) and the
acquisition period reached one millisecond, equivalent to 
the highest frequency of acquisition
of the Lab-PC+, without any problem. 

\bigskip
\bigskip
\textbf{3.2 xitris-client}
\bigskip
\bigskip

The xitris-client is the graphical interface between the user and the
xitris-server (is the client in the client-server model). We  use C with
GTK+ and GDK libraries to build the interfaces.  
The ADC board can be configured remotely using this interface. 
The xitris-client opens a socket to communicate with the
server, sends the ADC board configuration  and receives the data from the
server. The ADC board can be re-configured in real time. 
The client also saves and compress the raw data in the ``.xdat'' format 
using a sophisticated buffer that saves the data on the disk only when 
 a threshold value is reached. 
This buffer is very
efficient and provides a  fast implementation of the saving data process 
in real time, using limited resources.  

The .xdat format
takes advantage of the 12 bits length of the ADC data words, saving 4
measurements in a single integer variable of 32 bits. 
The buffer grows in the
memory dynamically. When the buffer is full, the data is saved on the hard
disk. A converter between .xdat and ASCII was also programmed. 
The .xdat is saved in the directory /yyyy/mm/dd where yyyy is the year, mm the 
month and dd the day when the data was acquired. 

The xitris-clients have two sub programs: ``ollin'' and ``tochtli''.  
Ollin is involved in the following processes: communicating with the server, 
displaying the data, and populate the buffer in real time. 
Tochtli is the interface with the user that controls Ollin software. 
The intercommunication between Ollin and Tochtli use an external variable via the memory. 
We use 36 register of 32 bits to intercommunicate
the configuration from the user to the Ollin (See Appendix A.2).

The buffer library is in the Ollin source code, while  the procedures to
create the graphics and the intercommunication with the server are 
in the ImageEngine library of the Ollin source code.

\bigskip
\bigskip
\textbf{3.3 xitris-www}
\bigskip
\bigskip

This module is responsible of plot and display the data on the
internet in real time. This software has two components: The real time
plotter (xoc) and the infrastructure for the website (public\_html). Both
codes work together in order to publish the plots and the raw data in real
time on the web in the following way: 
\begin{enumerate}
\item First, read the files generated by xitris-client (the .xdat files
  generated by Ollin and the .png files generated by xoc).
\item If the size of the .xdat archive changes, then it reads again the file 
and  generates the plot (making a .png file in a public path for the internet).
\item Finally, the code dynamically generate a web-page to display the plot in real time.
\end{enumerate}
The webpage of the database is generated in this module. We
includes the archives of each day of
observations.
The web site was tested with Apache tomcat 5 and 6. The source code is 
under the directory /public\_html. The Xoc code was written in C and works as a 
daemon of the system.  
Another web-pages can be included to introduce information about  the archives 
and the instrument.

\bigskip
\bigskip
\textbf{3.4 xitris-analyze}
\bigskip
\bigskip

This module can display the information  saved by xitris-client,
the plots are almost in real time, the time shift depends of the size of the
buffer of xitris-client.
It is possible to  
program algorithms to look for specific features over the signal
using a dedicated computer. 
We developed this module as log of xitris system. Each event or note can be
saved in this interface, the date and time is saved in the same buffer and can be recovered for further analysis.

\bigskip
\bigskip
\textbf{4. Results}
\bigskip
\bigskip

The source files of the four modules are packaged and can be downloaded from \cite{xitris}. 

\bigskip
\begin{center}
\includegraphics[width=1.0\textwidth]{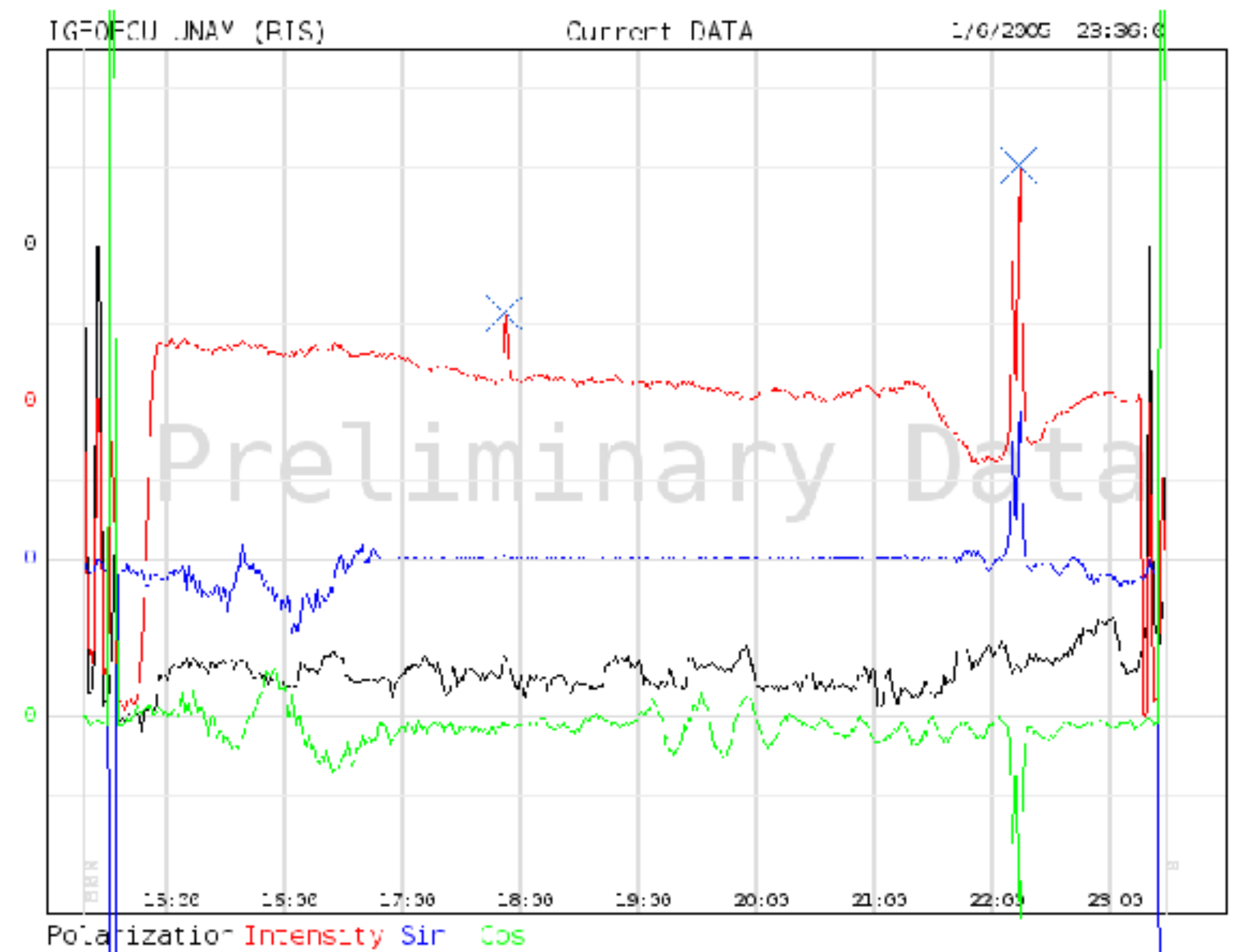}
\bigskip

\textbf{Figure 4} - Four channel plot published in internet in real time by xitris-www; Polarization vs Time (black), the Intensity (red), the Sin, and the Cos components of the signal (blue and green respectively). The cross shows solar events detected by the RIS.
\bigskip
\end{center}

\bigskip
\begin{center}
\includegraphics[width=1.0\textwidth]{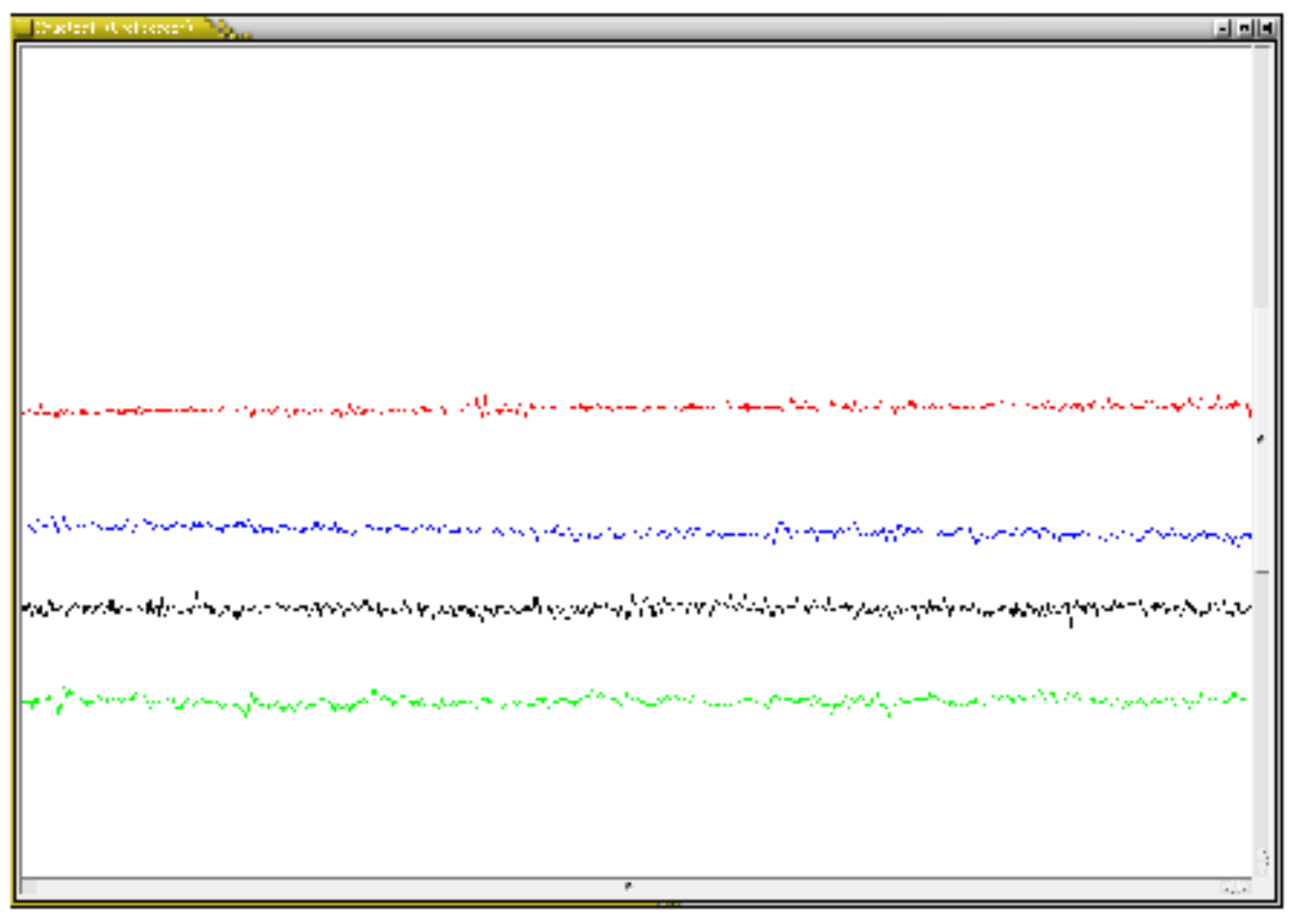}
\bigskip

\textbf{Figure 5} - Four channel plot from xitris-client (ollin interface) observed in the laboratory; this plot shows in real time the four signals from the radiotelescope.
\bigskip
\end{center}

\newpage
\begin{center}
\includegraphics[height=4cm]{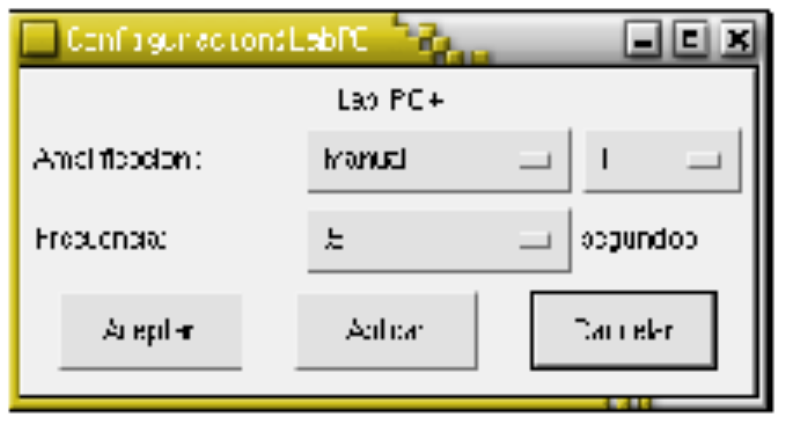}
\bigskip
\\
\textbf{Figure 6} - Interface for the ADC configuration (tochtli interface); we can configure the amplification and the sample frequency of the LabPC+ ADC board in real time. The changes also affect the ollin plot window.
\bigskip
\end{center}

\begin{center}
\includegraphics[width=1.0\textwidth]{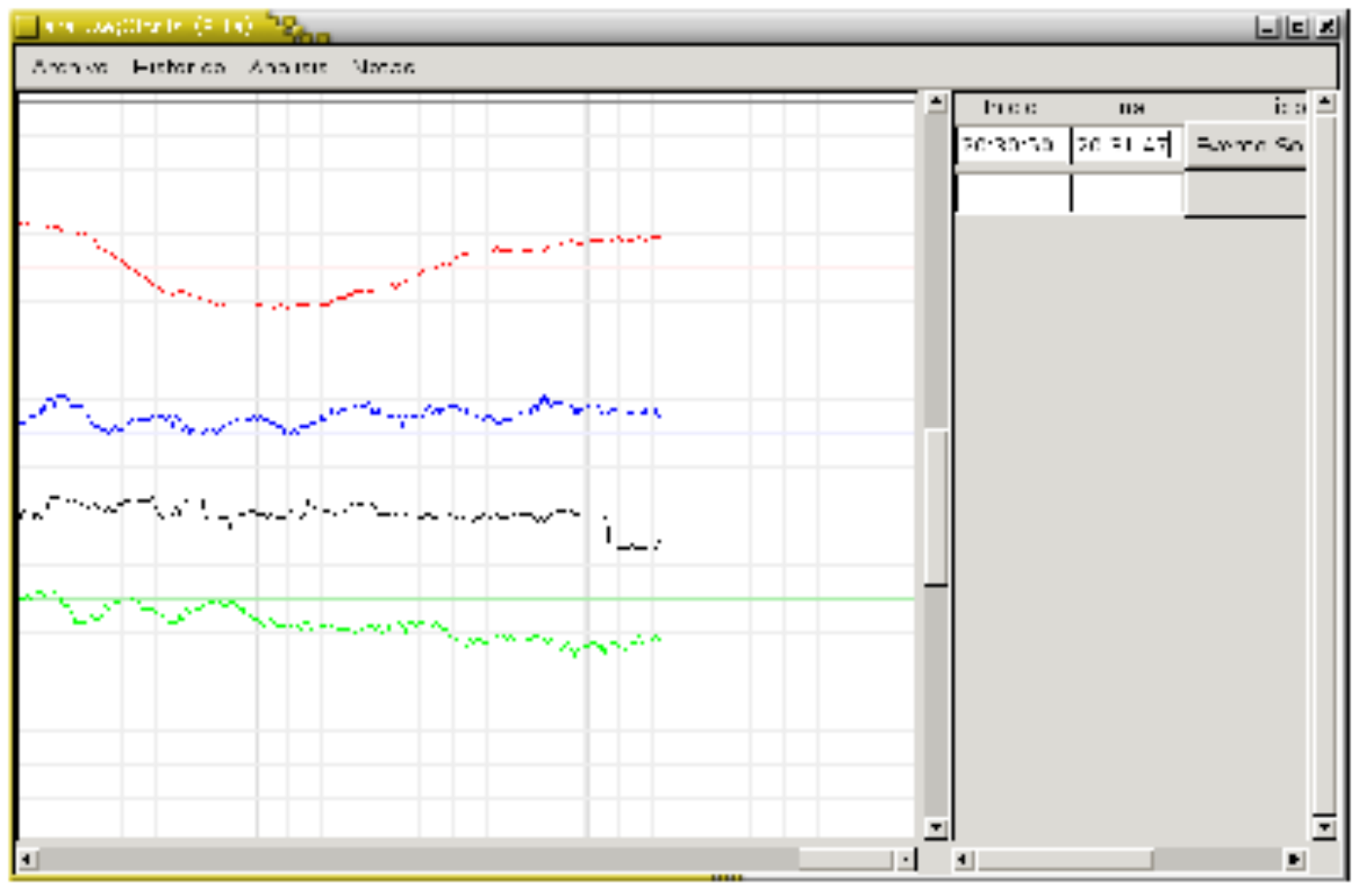}
\bigskip

\textbf{Figure 7} - Log of xitris (xitris-analyze); in this interface we can make annotations in real time over the signal, for example: in the case of malfunction of one component of the telescope, error in the pointing of the dish, no reference signal in the observation, the weather, a possible solar event, high noise, etc.
\bigskip
\end{center}

In Figures 4, 5, 6, and 7 we shows the interfaces of xitris that are running in the laboratory. The Figure 4 shows the plots that xitris publish in real time in internet via xitris-www, this plot is for full day of observations but is updated each 5 minutes. The Figure 5 shows the four channel plot that we running in the laboratory to observe the raw data of the signal in real time, the plot is updated in real time. The Figure 6 is the interface to configure the amplification and the clock sampling of the LabPC+ ADC board. The Figure 7 shows the log of the observation, in this interface we make annotations over the signal about problems or remarks due to external factors that could affect the observations.

Each interface can be running (and displayed) in different PCs. The information is carried trough the local network or internet inclusive, however the security for use public networks is not implemented yet.

The site of the RIS that use XITRIS can be founded in 
\begin{center}
http://cintli.geofisica.unam.mx/
\end{center}
Until now, there are four implementations of xitris-www in different observatories:
\begin{itemize}
\item The RIS: The Solar Radio Interferometer in Mexico City, Mexico. 
\item The Mexart: The Mexican Array Telescope in Morelia, Mexico.
\item The RT5: The Radio Telescope of 5 meters in Puebla, Mexico.
\item The IGA: The Institute of Geophysics in La Habana, Cuba.
\end{itemize}
\bigskip
\bigskip
\textbf{5. Conclusions}
\bigskip
\bigskip

The Xitris code was first released in 2005, since then the
code has been working  without important interruptions in the RIS facilities. 
This shows that Xitris is very stable and  reliable. 

The  distributed design of the code makes possible its adaptation to  
very restricted hardware infrastructure (old computers) and the C 
language is the key for a long life of the code. 

On the way, we developed a useful library for the ADC Lab-PC+ board 
of National Instruments with a GNU/GPL License 
that works in the GNU/Linux environments
with a high performance.

The eXtreme Programming was a very useful tool for dynamic design in a small
groups of programmers.

We have installed the code in four facilities. Actually, Xitris is the 
responsible of
deliver the data from the RIS to the Virtual Earth Solar Observatory (VESO) in
Mexico. 

Finally, Xitris can be expanded to incorporate  new ADC boards and new 
environments that involves ADC boards on remote control.

{\bf ACKNOWLEDGMENTS: The Debian user list to the invaluable help in several
  technical topics.}

\bigskip
\bigskip
\textbf{Appendix A.1: Conversion between 12 bits data and A2 32 bits integer in C}
\bigskip
\bigskip

In particular, the Lab-PC+ output consists on two consecutive bytes (of 8
bits). 
Therefore, it is necessary to  read two times the output register 
in order to get the  12  bits full data. 
The first read byte contains the 8 less significant bits, 
then the second read byte gives the 8
high significant bits (only use the last 4 bits). 
If the first bit of the second byte is zero, then the value 
is positive  
$$(0100 00000001)_2 = 2049_{10},$$
if this bit is one, then the value is negative
$$(1011 11111111)_2 = (0100 00000000)_2 + 1_2 = (0100 00000001)_2 = -2049_{10}.$$

In LabPC.c \cite{xitris}  we can found the implementation of the conversion between 12 bits data and A2 32 bits integer in C:
\begin{verbatim}
int ADFIFOregister(void){
  int lec1,lec2;
  int res; 
  int MASK = 128;
  int MASKCLEAR = 0x000000FF;
  if ((0x004 & inb(BASEPORT)) >> 2){ //Error Overflow!!
    return 5000;   //dummy flag
  }else if ((0x002 & inb(BASEPORT)) >> 1){ //Error Overrrun!!
    return -5000; //dummy flag
  }else{
    lec1= (MASKCLEAR & inb(BASEPORT+0x000A)); //byte low
    lec2= (MASKCLEAR & inb(BASEPORT+0x000A)); //byte high
    if ( (lec2 & MASK) == 128){ //negative value
      lec2 = (0xFFFFF000 | (lec2 << 8));
    }else{
      lec2 = (lec2 << 8); //positive
    }
    res = (lec1 | lec2);
    return res;
  }
}
\end{verbatim}

\bigskip
\bigskip
\textbf{Appendix A.2: Shared Memory}
\bigskip
\bigskip

In the main.c code of Tochtli \cite{xitris} is the definition to create the shared memory \cite{garcía1993unix}:
\begin{verbatim}
  int *Memoria = NULL;
  key_t Clave;
...
  Clave = ftok("/bin/ls",33);
  if (Clave == -1){
    printf("No memory key\n");
    exit(0);
  }
  Id_Memoria = shmget(Clave,sizeof(int)*36, 0777 | IPC_CREAT); /*create mem*/
  if (Id_Memoria == -1){
    printf("No Id\n");
    exit(0);
  }
  Memoria = (int*)shmat(Id_Memoria,(char*)0,0);
  if (Memoria == NULL){
    printf("No shared memory\n");
    exit(0);
  }
\end{verbatim}

In the main.c code of Ollin \cite{xitris} is the definition to read the shared memory  \cite{garcía1993unix}:
\begin{verbatim}
  Clave = ftok("/bin/ls",33);
  if (Clave == -1){
    printf("No Key\n");
    exit(0);
  }
  Id_Memoria = shmget(Clave,sizeof(int)*36,0777);
  if (Id_Memoria == -1){
    printf("No Id\n");
    exit(0);
  }
  Memoria = (int*)shmat(Id_Memoria,(char*)0,0);
  if (Memoria == NULL){
    printf("No Shared Memory\n");
    exit(0);
  }  
\end{verbatim}

\bibliography{xitris}

\begin{thebibliography}{1}

\bibitem{beck2000extreme}
K.~Beck.
\newblock {\em Extreme Programming Explained: Embrace Change}.
\newblock The XP Series. Addison-Wesley, 2000.

\bibitem{xitris}
Victor De~la Luz.
\newblock Xquetzal xitris.
\newblock http://sourceforge.net/projects/xquetzal/, April 2014.

\bibitem{garcía1993unix}
F.M.M. Garc{\'\i}a.
\newblock {\em Unix programaci{\'o}n avanzada}.
\newblock Ra-ma, 1993.

\bibitem{hall2000core}
M.~Hall.
\newblock {\em Core servlets and JavaServer Pages:}.
\newblock Prentice Hall PTR Core series. Prentice Hall PTR, 2000.

\bibitem{harlow1999developing}
E.~Harlow.
\newblock {\em Developing Linux Applications with Gtk+ and Gdk:}.
\newblock The Landmark Series. New Riders, 1999.

\bibitem{kochan1989unix}
S.G. Kochan and P.H. Wood.
\newblock {\em UNIX networking}.
\newblock Hayden Books UNIX system library. Hayden Books, 1989.

\bibitem{nissanke1997realtime}
N.~Nissanke.
\newblock {\em Realtime systems}.
\newblock Prentice Hall international series in computer science. Prentice
  Hall, 1997.

\bibitem{travis2007labview}
J.~Travis and J.~Kring.
\newblock {\em LabView for Everyone: Graphical Programming Made Easy and Fun}.
\newblock National Instruments Virtual Instrumentation Series. Prentice Hall,
  2007.

\bibitem{van1990data}
C.J. Van~Wyk.
\newblock {\em Data structures and C programs}.
\newblock Addison-Wesley series in computer science. Addison-Wesley Pub. Co.,
  1990.

\end{thebibliography}
\end{document}